# In situ mapping of indentation-induced densification and cracking in vitreous silica by nanofocus X-ray scattering

Johan F. S. Christensen[1], M. Faizal Ussama Jalaludeen[1], Søren S. Sørensen[1], Anders K. R. Christensen[1], Xuan Ge[1], Tao Du[1], Yuanzheng Yue[1], Anton Davydok[2], Christina Krywka[2], Lothar Wondraczek[3], Henning F. Poulsen[4], Morten M. Smedskjaer[1,*]

[1]Department of Chemistry and Bioscience, Aalborg University, Aalborg, Denmark
[2]Institute of Materials Physics, Helmholtz-Zentrum Hereon, Outstation at DESY, Hamburg, Germany
[3]Otto Schott Institute of Materials Research, University of Jena, Jena, Germany
[4]Department of Physics, Technical University of Denmark, 2800 Kongens Lyngby, Denmark
*Corresponding author. E-mail: mos@bio.aau.dk

**Abstract**
The practical strength of oxide glasses is greatly reduced by surface flaws that form during processing and use. Instrumented indentation can mimic such real-life damage events and induce flaws and cracking under controlled conditions. At the same time, instrumented indentation allows for systematic examination of the deformation and structural changes of the regions of the glass being indented. However, structural probing is nearly always performed *after* rather than *during* the sharp contact event, limiting our understanding of the indentation process. To overcome this, we here demonstrate the use of nanofocus X-ray scattering experiments to probe the local mechanical and structural response of vitreous silica during indentation. Two-dimensional mapping of the scattering pattern in the zone below a sharp diamond wedge indenter reveals local changes in the atomic structure and density as well as cracking behavior. These *in situ* experiments during indentation reveal the formation and evolution of the densification zone and cracking with nanoscale resolution. Understanding the interplay between structural densification and cracking behavior in glasses is deepened through this work, which is crucial for the development of more damage-resistant and thus stronger glasses as well as fundamental understanding of glass deformation mechanisms.

# Introduction

The mechanical reliability of oxide glasses is a major limitation for their use in many applications, as they are prone to fracture. They generally have high intrinsic strength, but their practical strength is compromised by the presence of surface flaws that act as local stress concentrators, leading to early fracture (1). Even though the severity of surface flaws can be reduced at the manufacturing stage, e.g., through fire-polishing or etching, surface flaws can be generated during usage when the glass comes in contact with sharp and hard objects (2, 3). Hence, there is a critical need to understand the mechanical and structural response of oxide glasses to sharp contact loading events, which can be investigated by instrumented indentation experiments. Indeed, the surface damage induced from indentation experiments often resembles the damage observed on glasses after real-life use (4).

The deformation of oxide glasses during indentation involves multiple types of deformation occurring simultaneously, but their individual extent differs significantly between glasses (5–7). Elastic deformation occurs already at low stress, while high tensile stresses induce formation of cracks. Oxide glasses are macroscopically brittle, i.e., they fracture without undergoing large-scale plastic deformation, but they exhibit plastic deformation on the microscale, e.g., during indentation (8, 9). This results in an imprint on the surface, as locally high stresses of magnitude similar to the hardness of oxide glasses (~5 GPa) are reached. The plastic deformation can be divided into densification and shear flow, the latter being a volume-conserving process with the displacement of matter. Ultimately, a complex deformation zone often evolves during contact, potentially inducing different crack patterns during loading and/or unloading, depending on glass properties and contact loading conditions (10, 11).

Various studies have investigated the indentation-induced cracking behavior of oxide glasses, mainly by inspecting the crack patterns as well as other indentation-induced deformation and structural changes after the contact event (11–14). Mapping the surface topography can be used for quantifying the pile-up caused by shear flow and also the amount of densification (7, 15). Microscale plasticity dissipates stress, and especially densification has been proposed to reduce crack initiation (3, 16). The shape and extent of the densification zone formed below a sharp indenter have been investigated using different techniques, including Raman spectroscopy (17–21), Brillouin spectroscopy (22, 23), and chemical dissolution techniques (24, 25). Many studies have focused on vitreous silica, which has a high propensity for densification and only shows very limited shear flow upon indentation with a relatively blunt tip, such as Vickers or Berkovich. While it is known that the deformation mechanism transitions from densification to shear flow for sharper indenter geometries (3, 11), there are considerable differences in the reported densification profiles for vitreous silica even for similar indentation conditions (25).

Our current understanding of the indentation response of oxide glasses is limited by the fact that structural probing is typically performed *after* indentation, although *in situ* measurements and simulations indicate reversible indentation-induced deformation (26, 27). Hence, *in situ* measurements of stress, deformation, and cracking are important for advancing our understanding of surface damage formation, elucidating the underlying mechanisms, and ultimately for developing stronger glasses. The use of inverted microscopes, allowing for observation of the indented surface through transparent samples, has proven useful for *in situ* observation of crack formation (10, 26, 28). Stress fields have been investigated *in situ* by birefringence measurements (29, 30). However, these *in situ* experimental studies have not quantified the indentation-induced deformation zone in oxide glasses. On the other hand, modelling of indentation-induced deformation has also been performed using, e.g., finite element or peridynamics methods, but these are limited by the accuracy of the constitutive models used (20, 25, 31, 32).

To enable *in situ* observation of indentation-induced deformation, we here map the local mechanical and structural response of vitreous silica during indentation using nanofocus X-ray scattering. While related measurements have previously been attempted for different glass systems (33–35), we here report improved resolution based on recent advances in beamline optics and our developed experimental design. In the previous work, two-dimensional spatially resolved strains around a circular hole in a plate under compression (33) and in a thin sample under point indentation (34) were measured on bulk metallic glasses. The densification zone in vitreous silica resulting from wedge indentation was investigated in preliminary experiments, achieving a spatial resolution of 1 to 2 µm across a two-dimensional projection of the indented region (35). While these experiments aimed at evaluating the suitability of *in situ* X-ray scattering for exploring indentation-induced densification *per se*, they did not produce consistent results across different indents and provided only limited information regarding the shape of the deformation zone.

In this work, we measure the local X-ray structure factors $S(Q)$ of vitreous silica around the indentation site with a spatial resolution down to 500 nm using a nanofocused X-ray beam before, during, and after indentation. That is, by measuring the local X-ray scattering pattern up to scattering vector $Q \geq 5$ Å$^{-1}$ in the investigated area by 2D raster scanning (Fig. 1), we obtain the spatially resolved first sharp diffraction peak (FSDP) of $S(Q)$. From the FSDP position, we construct high-resolution maps of the indentation-induced densification zone at progressively increasing loads during indentation as well as after complete unloading. The local densities are evaluated using a previously established experimental relation between FSDP position and density for vitreous silica (36). Our results reveal significant differences in the shapes of the temporary and permanent densification zones and the related structural changes in the atomic network. We are also able to observe cracking *in situ* based on the intensity of the scattered signal.

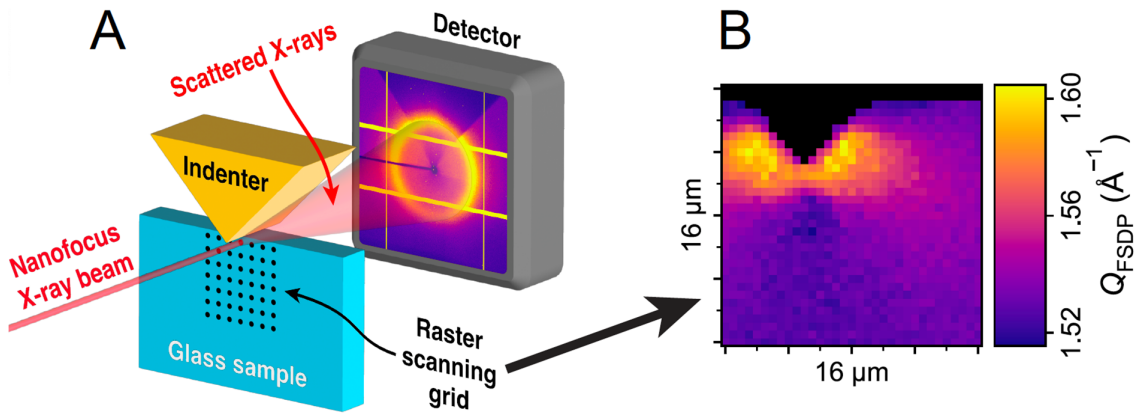

**Figure 1. (A)** Schematic of the experimental setup. Indentation is performed *in situ* on the vitreous silica sample using a diamond wedge indenter, while the local scattering pattern of the glass is obtained for each of the individual points in the measurement grid by raster scanning. The X-ray scattering measurements are performed in transmission mode, with the intensity of the scattered X-rays being measured with a 2D detector. **(B)** Example of 2D map generated by extracting the first sharp diffraction peak position ($Q_{FSDP}$) from each point in a raster scan, performed while applying a 4 N load.

## Results

***Indentation-induced changes in the FSDP.*** Upon applying indentation load on vitreous silica, large changes are observed in the X-ray structure factor $S(Q)$ as measured locally at the points in the two-dimensional scanning grid. As shown in Figs. 2A-B, the intensity of the FSDP, when measured at representative points near the indentation site, monotonically decreases with increasing indentation load. Upon unloading, the intensity of the FSDP partly recovers, but remains significantly lower than that of the pristine glass prior to indentation due to irreversible changes in the local structure. In addition to the change in intensity, the FSDP position ($Q_{FSDP}$) also shifts to higher $Q$ values when indentation load is applied and partially recovers after unloading. As described in the Supporting Information Text, we do not observe significant changes in $S(Q)$ in the areas far away from the indentation site upon repeated X-ray beam exposure, i.e., the observed changes are induced by the indentation process and not due to X-ray beam damage (*SI Appendix*, Fig. S1). To trace the local variation of the FSDP in the area around the indentation site during indentation, we construct two-dimensional maps of its position. This is done by extracting the local $Q_{FDSP}$ value from each scanned point, as shown in Fig. 1B for the glass under 4 N indentation load.

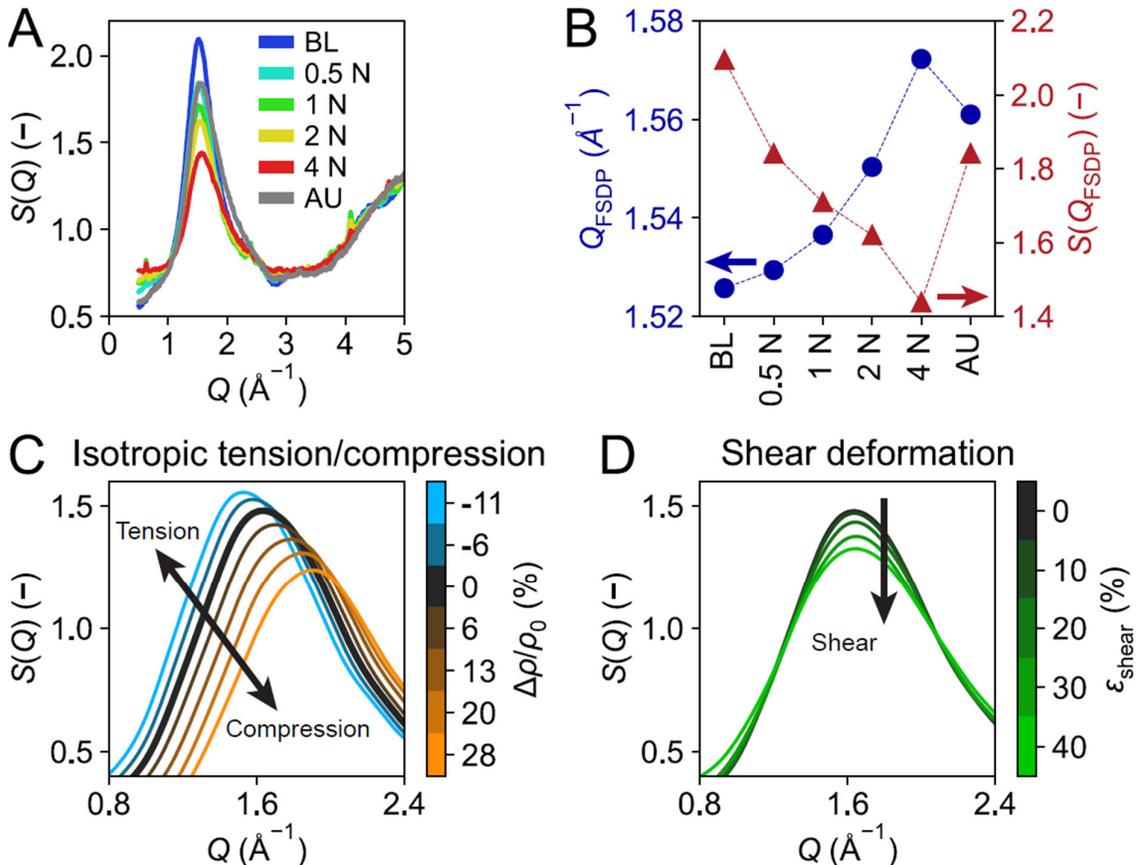

**Figure 2.** **(A)** Examples of measured X-ray structure factor, $S(Q)$, at points near the indentation site at each stage of the indentation cycle (BL = Before Loading, AU = After Unloading). The first sharp diffraction peak (FSDP) intensity monotonically decreases, while the FSDP position shifts slightly to higher $Q$ values during indentation loading, both of which recover partly after unloading. **(B)** FSDP position ($Q_{FSDP}$) and peak intensity ($S(Q_{FSDP})$) plotted as a function of the stage of the indentation cycle. **(C-D)** Evolution of the X-ray FSDP as determined from molecular dynamics simulations of vitreous silica upon (C) isotropic tension or compression and (D) shear deformation, showing that $Q_{FSDP}$ increases for compression, decreases for tensile deformation, and remains constant for shear deformation. $\Delta\rho/\rho_0$ is the relative change in density ($\rho$), and $\varepsilon_{strain}$ is the shear strain.

The FSDP of vitreous silica and other oxide glasses is related to their medium-range order. $Q_{FSDP}$ values of around 1.5 Å$^{-1}$ correspond to distances of around 4.2 Å, which can be considered as the mean interplanar spacing if considering the existence of a "quasilattice" (36, 37). The FSDP of silicate glasses has been shown to be linked to the ring size distribution, including resolving the FSDP into contributions from ring structures of different sizes and hence different $Q$ positions (38). Thus, changes of the FSDP are related to changes of the medium-range structure, and the shift of the FSDP to higher $Q$ values upon compression implies both a more densely packed structure and an increase in the fraction of smaller relative to larger rings.

*Mapping the densification zone.* The mapping of the local FSDP during indentation allows us to extract information about the local densification through a previously reported correlation between $Q_{FSDP}$ and the density of vitreous silica (36). The relation was established through X-ray scattering experiments on vitreous silica samples subjected to compression at 5 GPa and temperatures ranging from 500 to 800 °C. Tan and Arndt (36) also found that the intensity of the FSDP decreases with increasing density. Since indentation involves shear flow in addition to pure densification, we have performed molecular dynamics simulations (see Methods for details) of vitreous silica to evaluate how the FSDP is affected by different types of deformation. As shown in Fig. 2C, our simulations confirm that $Q_{FSDP}$ increases when the glass is isotropically compressed and the density increases. Similarly, $Q_{FSDP}$ decreases upon isotropic tensile deformation, leading to reduced density. Our simulations also capture the trend of the change in the intensity of the FSDP with both compressive and tensile deformation. When subjecting the simulated vitreous silica to pure shear deformation, we observe that the FSDP position remains unchanged, but its intensity decreases (Fig. 2D). *SI Appendix*, Figs. S2-S3 further show the evolution of the simulated $S(Q)$ upon deformation for a wider $Q$ range (up to 20 Å$^{-1}$), revealing that the structure factor is much less affected during shear deformation, compared to the changes induced by compressive or tensile deformation.

Overall, these results justify that the density can be evaluated from the FSDP position even when shear deformation occurs. In detail, we calculate the local density ($\rho$, in units of g cm$^{-3}$) of vitreous silica by the relation (for $Q_{FSDP}$ in units of Å$^{-1}$) (36),

$$\rho = 1.621 \text{ g cm}^{-3} + 41.9 \text{ g cm}^{-3} \text{ Å}^3 \, (Q_{FSDP}/2\pi)^3 \,. \tag{1}$$

By estimating the local density from the spatially resolved measurements of $S(Q)$, maps of the density in the area around the indentation site throughout the indentation process have been generated (Fig. 3). First, we note that the glass appears fully homogenous before indentation. Upon contact, a zone of higher density appears near the contact point with the indenter (see indenter silhouette included in Fig. 3), with the extent of densification monotonically increasing when a higher indentation load is applied. Upon unloading, i.e., after the indenter has been completely removed, a semicircular zone of permanent densification is observed. This permanent densification zone after unloading is much smaller than the temporarily induced densification zone formed during indentation, but the magnitude of the local densification just around the indentation site is similar to that induced by 4 N load.

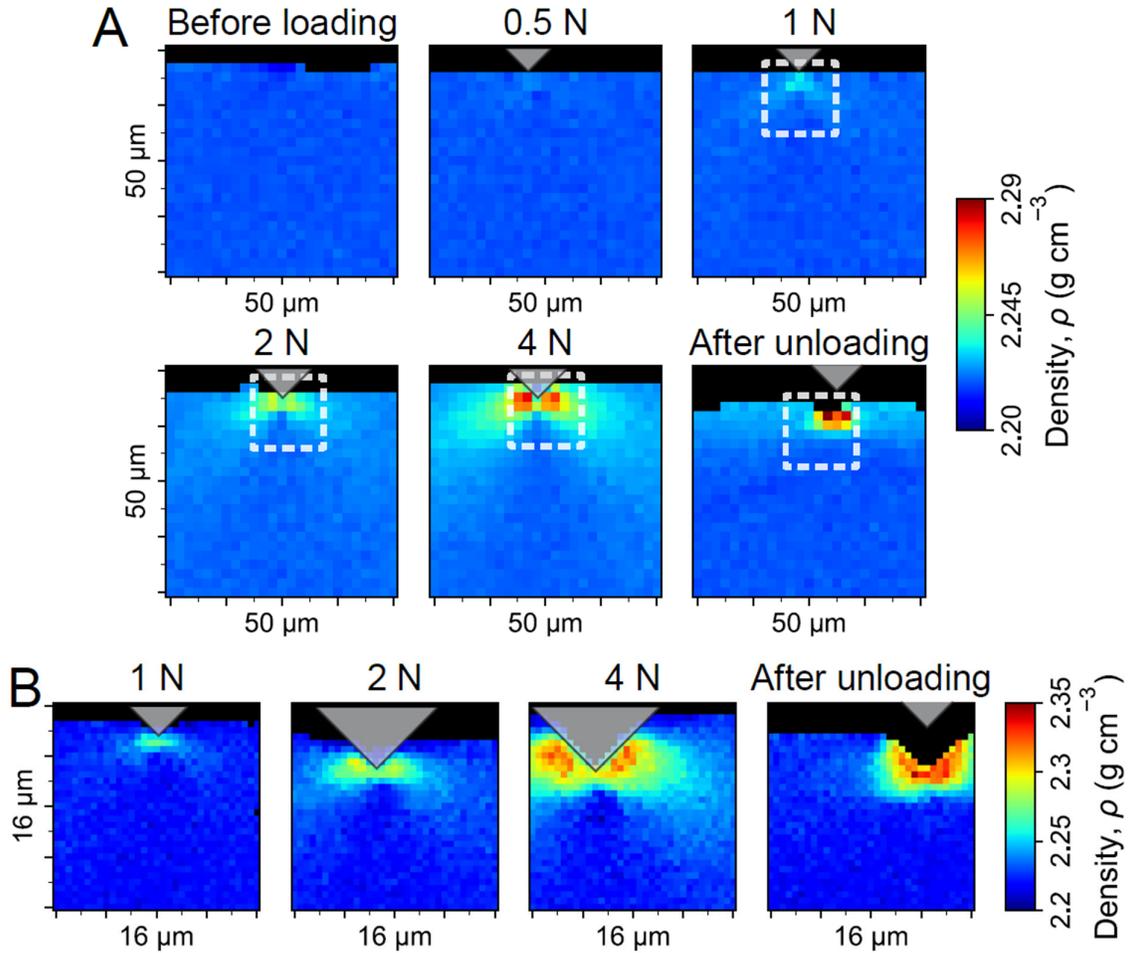

**Figure 3.** Heat maps of the estimated local density in vitreous silica at different stages of the indentation cycle for **(A)** large-area scans of 2 µm resolution and **(B)** small-area scans of 0.5 µm resolution. The white-dashed marked boxes in panel A denote the areas that have been subjected to the additional scans shown in panel B. The color scales differ between panels A and B to improve the visibility of different features (*SI Appendix*, Fig. S4 shows all panels using the same color scale). The grey triangles represent the position of the indenter, which is slightly shifted between the scans.

*Indentation-induced cracking.* In addition to the density maps estimated from the FSDP position, we have also generated maps of the maximum intensity of the FSDP (Fig. 4). The latter is related to the extent of interaction between glass and X-rays, thus revealing where less glass is present, i.e., cracks in the sample, in addition to the shape of the indentation imprint. At 2 N load, a median crack is observed, extending downwards from the bottom of the indentation imprint. As the load increases to 4 N, the median crack opens more, making it more visible. Upon unloading, the median crack has closed and can hardly be observed, while lateral cracks starting from below the indentation imprint and extending sideways appear. We note that densification of the glass also reduces the FSDP intensity (Figs. 2A-C) (36), i.e., the zones of increased density (Fig. 3) cause corresponding zones of lower peak intensity in Fig. 4. In other words, the semicircular zones of lower FSDP intensity just around the indentation site (Fig. 4B) are caused by this effect and not displacement of material. Also, we note that the peak intensity does not reach zero in the cracks, but this can be caused by the crack being narrower than the size of the X-ray beam or potential twist of the crack when it propagates through the sample. Thus, the method is only applicable for observing cracks of a significant width, i.e., the crack width should at least be comparable to the X-ray beam size.

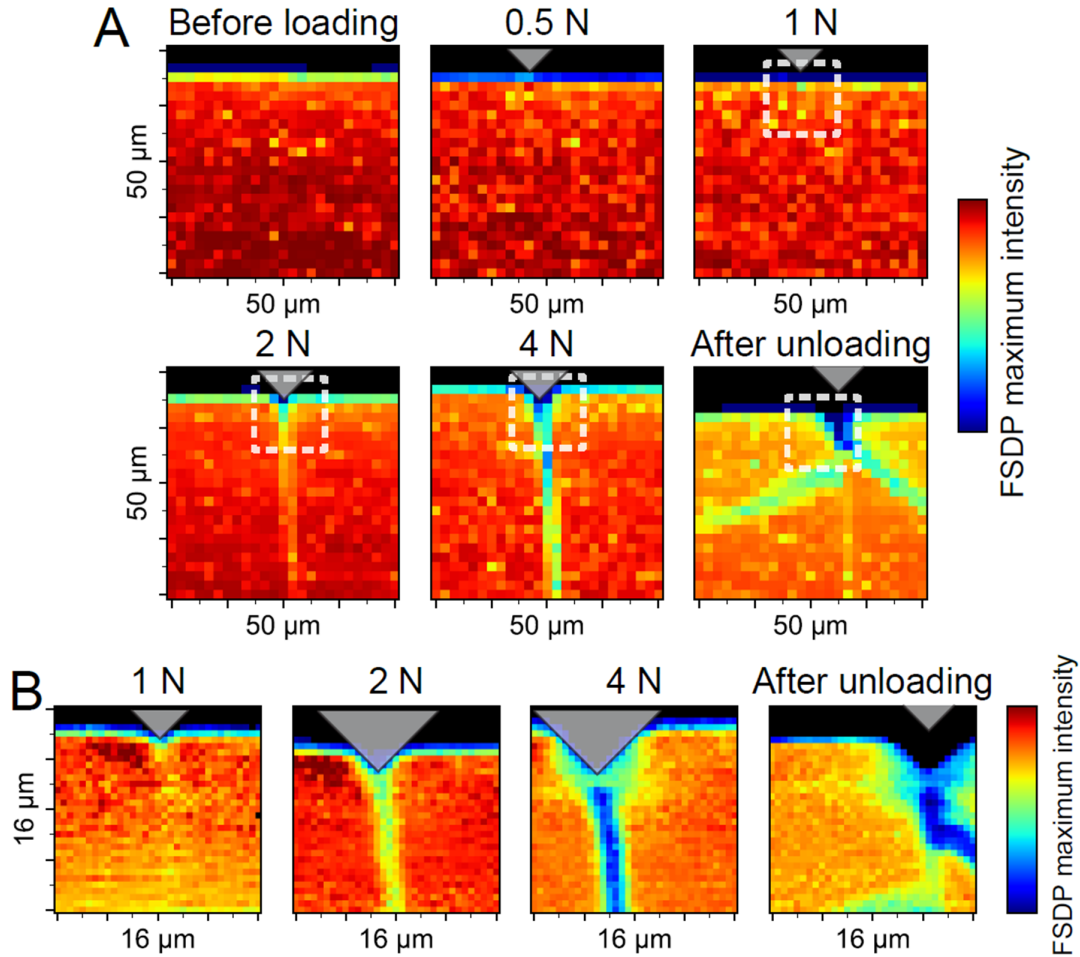

**Figure 4.** Heat maps of the intensity of the first sharp diffraction peak FSDP (instead of the estimated density in Fig. 3) at different stages of the indentation cycle for **(A)** large-area scans of 2 μm resolution and **(B)** small-area scans of 0.5 μm resolution. The white-dashed marked boxes in panel A denote the areas that have been subjected to the additional scans shown in panel B. The FSDP intensity correlates with the amount of glass present, i.e., the indentation imprints are visible as well as any formed cracks. The grey triangles represent the position of the indenter, which is slightly shifted between the scans.

As cracking allows for release of stress, it affects the stress field and hence the deformation zone. For example, this is apparent from the "butterfly-shape" of the densification zone seen at loads of 2 and 4 N in Fig. 3, where only limited densification is observed below the indenter tip. This is likely related to the formation of the median crack, which concentrates stresses in a different manner compared to the non-cracked glass. Such cracking can be problematic, e.g., if aiming to map the densification zone without the complexity of crack formation and the associated stress release (25). Blunter indenters or lower loads could be used to mitigate crack formation. However, cracking is an important aspect in relation to the mechanical performance of oxide glasses, and the present results show that the nanofocus X-ray scattering technique can be used for samples containing cracks. This allows for the study of various types of deformation, including damage formation in real-life scenarios.

## Discussion

We have shown how nanofocus X-ray scattering enables *in situ* high-resolution two-dimensional mapping of the indentation-induced local atomic-scale structure of oxide glasses during indentation. Upon indentation of vitreous silica, the spatially resolved scattering patterns enable *in situ* evaluation of the local density, as well as glass displacement, revealing the indentation imprint and cracking. Notably, the nanofocus X-ray scattering technique also captures the shape of the indentation imprint *in situ*, thereby allowing for determination of the *in situ* hardness. Employing an indentation procedure with several load steps, the progressive development of the investigated features can be followed during the indentation sequence, providing important insight into the indentation response of vitreous silica and the underlying deformation mechanisms. Such direct observation of the deformation zone *in situ* during indentation can be used to develop and calibrate constitutive models for glass deformation used in finite element and peridynamics simulations. In addition, the *in situ* visualization of indentation can help explain anomalous behaviors of vitreous silica, for example, its anomalous softening upon compression, which has not been fully understood (39).

The indentation imprints produced on glasses are generally known to be of shallower depth than expected from the indenter geometry due to elastic recovery (40). Indeed, the generated images directly reveal that the indentation imprint shape follows that of the indenter (90° angle) during loading, while the elastic recovery results in a larger imprint angle of ~105° after unloading (see Figs. 3B and 4B). The permanent deformation of vitreous silica by indentation is known to predominantly involve densification and only involve limited deformation through shear flow. This is typically observed in the case of a Vickers indenter tip, but we note that a sharper indenter tip has been used herein. In turn, this shifts the deformation mechanism from densification to more shear flow (3, 11). No pile-up of material beside the indentation imprint has been observed in the present experiments, i.e., no significant shear flow is observed for silica glass despite the sharpness of the used indenter. However, an observable pile-up would be expected for glasses with a higher propensity for shear flow compared to vitreous silica.

In comparison to the earlier study of Fuhrmann et al. (35), who also probed the indentation-induced densification zone in vitreous silica, we here observe the densification zone in greater detail for a series of indentation loads. To achieve this, first, we scan at an increased spatial resolution (0.5 vs. 1-2 µm), enabled by a smaller beam spot size (300×300 vs. 800×800 nm$^2$). Second, a more pronounced densification zone is induced by using a blunter indenter (90° vs. 60° angle), shifting the deformation mechanism more towards densification (11). Also, more deformation is induced as we use a higher indentation load (up to 4 vs. ~2.5 N) and a thinner sample (150 vs. 467 µm), increasing the load per contact length by up to a factor of ~5. Ultimately, these improvements enabled the *in situ* observation of the gradual evolution of the densification zone and cracking during the indentation experiment. The showcased technique also enables the study of the directional deformation of the sample during indentation, as this information is encoded into the azimuthal angle of the scattered X-rays (33). This has, however, not been attempted in the present study due to the signal-to-noise ratio and shadows on the detector, both originating from the experimental design, i.e., only non-directional densification has been evaluated. In detail, by performing the integration of the detector data in sections over the azimuthal angle, the shift of peaks in $S(Q)$, such as $Q_{FSDP}$, can be deconvoluted into three strains. That is, into two perpendicular normal strains in the plane perpendicular to the beam and the shear strain in the same plane. Furthermore, the strain components in the third dimension can in principle be obtained through measurements at different angles between the sample and X-ray beam.

Overall, our study has demonstrated the capabilities of the employed method for studying the response of various glasses to widely varying indentation conditions. This will advance our understanding of the contact mechanics of glasses and provide important data for the development of models for glass deformation, in turn helping the design of stronger glasses with smaller propensity to cracking.

## Materials and Methods

*X-ray synchrotron experiments.* Vitreous silica (Corning® High-Purity Fused Silica HPFS® 7980, OH content of 800-1000 ppm (41)) was cut, ground, and polished to make a sample of rectangular cuboid geometry with a thickness of 150 μm. All faces relevant for the indentation and X-ray scattering experiments were polished to an optical finish using water-free diamond suspensions.

X-ray nanofocus scattering experiments were performed at the Nanofocus Endstation of the MiNaXS (P03) beamline at PETRA III, DESY, Hamburg (42). A highly focused monochromatic X-ray beam (19.7 keV) with a beam spot size of either 300×300 or 1500×1500 nm$^2$ was used. The spot size refers to the full width at half maximum of the X-ray beam intensity, and the flux of the beams was approximately $10^9$ s$^{-1}$ and $10^{10}$ s$^{-1}$, respectively, for the smallest and largest beam. The synchrotron beam was focused down by means of KB mirrors installed directly before the sample, at a distance of 90 m from the source. The focal depth of the mirrors was 50 μm, with the beam size increasing by 1 μm at each millimeter distance from the focal spot. The central part of the sample was placed at the focal plane with the help of an optical microscope. The experiments were performed in transmission mode (through the sample of 150 μm thickness), with the scattered X-rays being detected by a DECTRIS EIGER S 9M 2D detector placed at a sample-to-detector distance of 210 mm (Fig. 1A). As a result, we detect the scattered X-rays up to $Q$ of around 5-6 Å$^{-1}$, making it infeasible to convert the obtained X-ray structure factor $S(Q)$ data to real-space pair distribution function data. $Q$ is the scattering vector defined as $4\pi\sin(\theta)\lambda^{-1}$, with $\theta$ being half the $2\theta$ scattering angle and $\lambda$ the incident wavelength. The experiment geometry and exact sample-to-detector distance were calibrated using lanthanum hexaboride (LaB$_6$) powder as the standard.

Each X-ray scattering measurement at a local point on the sample yielded a 2D detector image. An exposure time of 5 and 1 s was used for the beams of 300×300 and 1500×1500 nm$^2$ spot size, respectively. After masking the detector image areas affected by the shadows from the beam stop, indenter, and sample holder, azimuthal integration of the data was performed with pyFAI (43), yielding the intensities of the scattered X-rays over the measured $Q$ range. The intensities were converted into $S(Q)$ with the PDFgetX3 software (44) using the known composition of the sample (SiO$_2$) and background correction with the measured air scattering. The conversion to $S(Q)$ divides the scattering into self-scattering and distinct-scattering contributions, the latter containing all structural information, which is here presented as $S(Q)$ (45). Two-dimensional mapping of the glass' local $S(Q)$ was achieved by moving the sample (together with the complete indentation setup described below) in the plane perpendicular to the beam. When using the X-ray beam of 300×300 nm$^2$ spot size, raster scans on 16×16 μm$^2$ areas were obtained by moving the sample relative to the beam in steps of 0.5 μm. With the 1500×1500 nm$^2$ beam, areas of 50×50 μm$^2$ were scanned using a step size of 2 μm.

*In situ* indentation on the sample was performed using the custom-built indentation setup described elsewhere (46) and shown in *SI Appendix*, Figs. S5 and S6 (see also sketch in Fig. 1A). The setup was equipped with a diamond wedge indenter of 90° included angle and 500 μm wedge length (Synton-MDP). Wedge indentation across the whole thickness of the sample was preferred over the traditional point indentation, such as Vickers indentation, as a wedge creates uniform stress and deformation zones across the thickness of the sample when not considering edge effects. Thus, the contact edge of the indenter wedge was aligned parallel to the top sample surface, as well as to the X-ray nanobeam, prior to indentation. Hence, when the X-ray nanobeam goes through the 150 µm thick sample, the stress and deformation of the sample are assumed to be uniform. Any misalignment would cause the probed volumes to be inhomogeneous across the sample thickness, yielding scattering signals of mixed sample states. Therefore, careful alignment was performed utilizing the indentation setup's capability of rotating the whole setup, as well as independently rotating the sample, around all axes.

Indentation was then performed by bringing the sample in contact with the wedge indenter, and increasing the load in steps to 0.5, 1, 2, and 4 N, followed by complete unloading. Raster scans were performed on the sample at the indentation site just before contact, during the hold time at each load, as well as after unloading. Scans with the beam of $1500 \times 1500$ nm$^2$ spot size were made at all steps, while scans with the $300 \times 300$ nm$^2$ beam were performed at loads of 1, 2, and 4 N and after unloading.

*Molecular dynamics simulations.* To understand the effect of compression, tension, and shear deformation on $S(Q)$ and thus guide the interpretation of the experimental data, we performed classical molecular dynamics simulations of vitreous silica. The initial system consisted of 8100 atoms, which was subjected to a melt-quenching procedure to prepare a cubic vitreous silica structure. The interatomic interactions were described using the BKS potential (47), and the long-range Coulombic interactions were truncated using the Wolf shift method. The system was first melted in the *NPT* ensemble at 5000 K and zero pressure for 45 ps and then quenched to 300 K at 1 K ps$^{-1}$. The resulting structure was further equilibrated at 300 K and zero pressure for 70 ps to obtain the glass structure. Simulated deformation of the prepared vitreous silica was performed at 300 K to obtain the X-ray structure factor $S(Q)$ and, in turn, FSDP position and intensity at different isotropic tension and compression states, as well as pure shear states. All simulations were performed using LAMMPS software (48).

## Acknowledgments

We acknowledge Sidsel M. Johansen (Aalborg University) for experimental assistance. This work was supported by the European Union (ERC, NewGLASS, 101044664) and the ESS lighthouse on hard materials in 3D, SOLID, funded by the Danish Agency for Science and Higher Education (8144-00002). We acknowledge DESY (Hamburg, Germany), a member of the Helmholtz Association HGF, for the provision of experimental facilities. Parts of this research were carried out at PETRA III. Data was collected using the Nanofocus Endstation of the MiNaXS (P03) beamline operated by Helmholtz-Zentrum Hereon. Beamtime was allocated for proposal I-20230040. We also acknowledge the Danish Agency for Science, Technology, and Innovation for travel funding through DanScatt.

# Supporting Information

*for*

# In situ mapping of indentation-induced densification and cracking in vitreous silica by nanofocus X-ray scattering

Johan F. S. Christensen[1], M. Faizal Ussama Jalaludeen[1], Søren S. Sørensen[1], Anders K. R. Christensen[1], Xuan Ge[1], Tao Du[1], Yuanzheng Yue[1], Anton Davydok[2], Christina Krywka[2], Lothar Wondraczek[3], Henning F. Poulsen[4], Morten M. Smedskjaer[1,*]

[1]*Department of Chemistry and Bioscience, Aalborg University, Aalborg, Denmark*
[2]*Institute of Materials Physics, Helmholtz-Zentrum Hereon, Outstation at DESY, Hamburg, Germany*
[3]*Otto Schott Institute of Materials Research, University of Jena, Jena, Germany*
[4]*Department of Physics, Technical University of Denmark, 2800 Kongens Lyngby, Denmark*
[*]*Corresponding author. E-mail: mos@bio.aau.dk*

## Supporting Information Text

***Investigation of potential beam damage.*** The intense X-ray radiation used in the nanofocus experiments can potentially induce changes in the glass structure, known as beam damage (1). To test for the presence of any beam damage, we compare the obtained X-ray structure factors after the same area of the glass has been repeatedly scanned during the experiment. To minimize the effect of the indentation deformation and thus isolate the potential effect of beam damage, we here compare structure factors obtained at two approximate points relatively far from the indentation site, as shown in Fig. S1. The glass at approximately 35 μm below the indentation site is subjected to X-ray exposure only during the scans with the 1500×1500 $nm^2$ beam. Approximately 10 μm below the indentation site, the sample is also exposed to X-rays during the 300×300 $nm^2$ beam scans. For the latter scans, five times longer exposure time is used, further increasing the risk of beam damage.

For both investigated points, we find no changes in the X-ray structure factor when comparing the initial measurement with that after unloading, indicating the absence of beam damage. However, a small, gradual decrease in the first sharp diffraction peak (FSDP) intensity is seen when load is applied to the sample, but the observed change is reversible and disappears after unloading. This is barely visible farthest from the surface (Fig. S1A) but can clearly be seen in the area of the high-resolution scans (Fig. S1B). As the change is reversible and much more pronounced closer to the indentation site, it is ascribed to be an effect of the indentation deformation. We note that the change is visible (but small) at 1 N load in Fig. S1B, whereas it is not visible at 1 N load in Fig. S1A. As no scans with the 300×300 $nm^2$ beam were performed prior to the 1 N scan, this further indicates that the effect is caused by the indentation. Overall, we therefore conclude that no significant X-ray beam damage has occurred during repetitive mapping of the same area in the present experiments on vitreous silica.

### *Reference*
1. A. Martinelli, *et al.*, Reaching the Yield Point of a Glass During X-Ray Irradiation. *Phys Rev X* **13**, 041031 (2023).

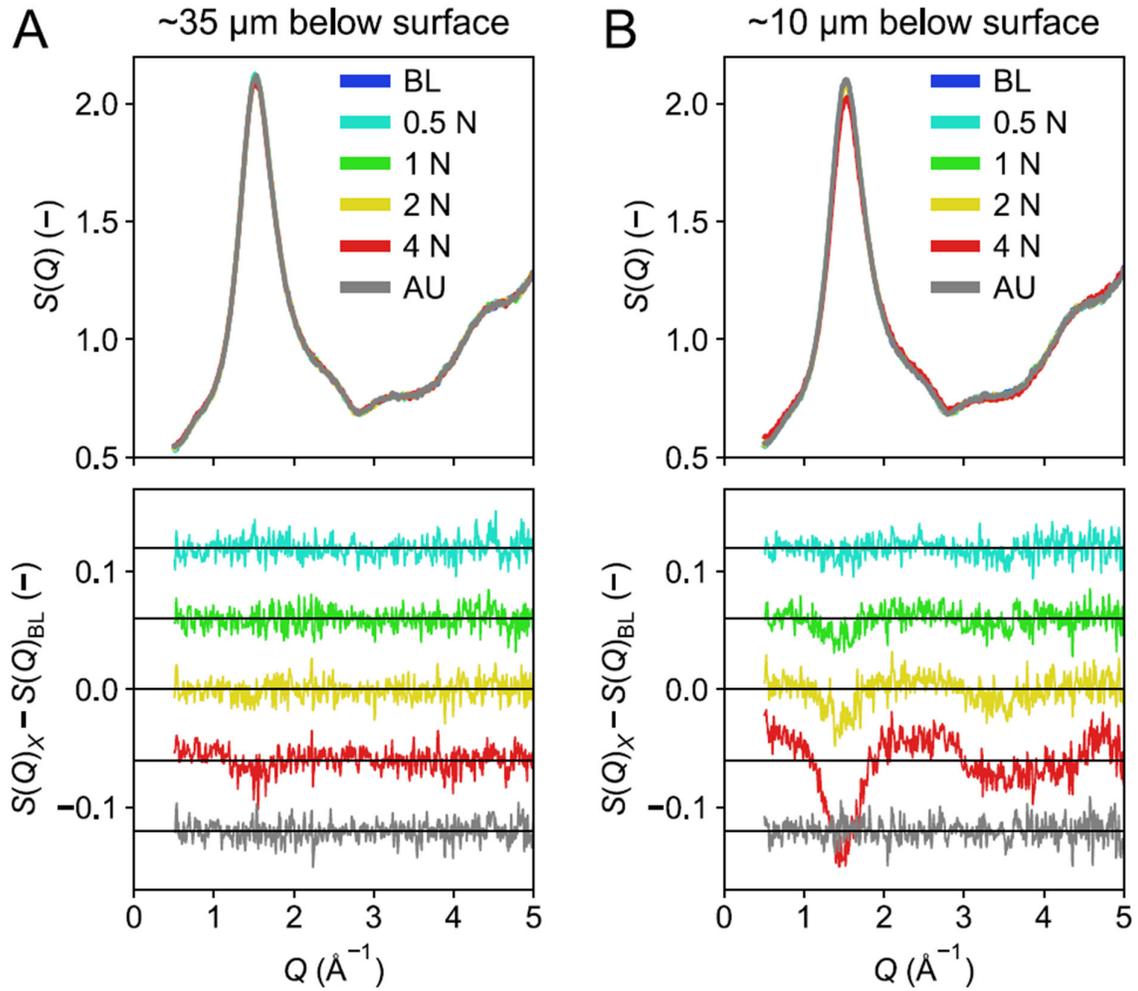

**Fig. S1.** Investigation of potential beam damage, i.e., X-ray-induced changes, by comparing the obtained X-ray structure factors $S(Q)$ when the sample is scanned repeatedly at the different steps in the indentation sequence. Scans are also performed before loading (BL) and after unloading (AU). Two approximate points located at **(A)** ~35 μm and **(B)** ~10 μm, respectively, below the indentation site (when measured from the original glass surface) are investigated here. The first point (panel A) is only subjected to scans with the 1500×1500 nm$^2$ beam, while the second point (panel B) is also scanned with the 300×300 nm$^2$ beam. The lower panels show the difference plots $S(Q)_X - S(Q)_{BL}$ ($X$ = 0.5 N, 1 N, 2 N, 4 N, AL).

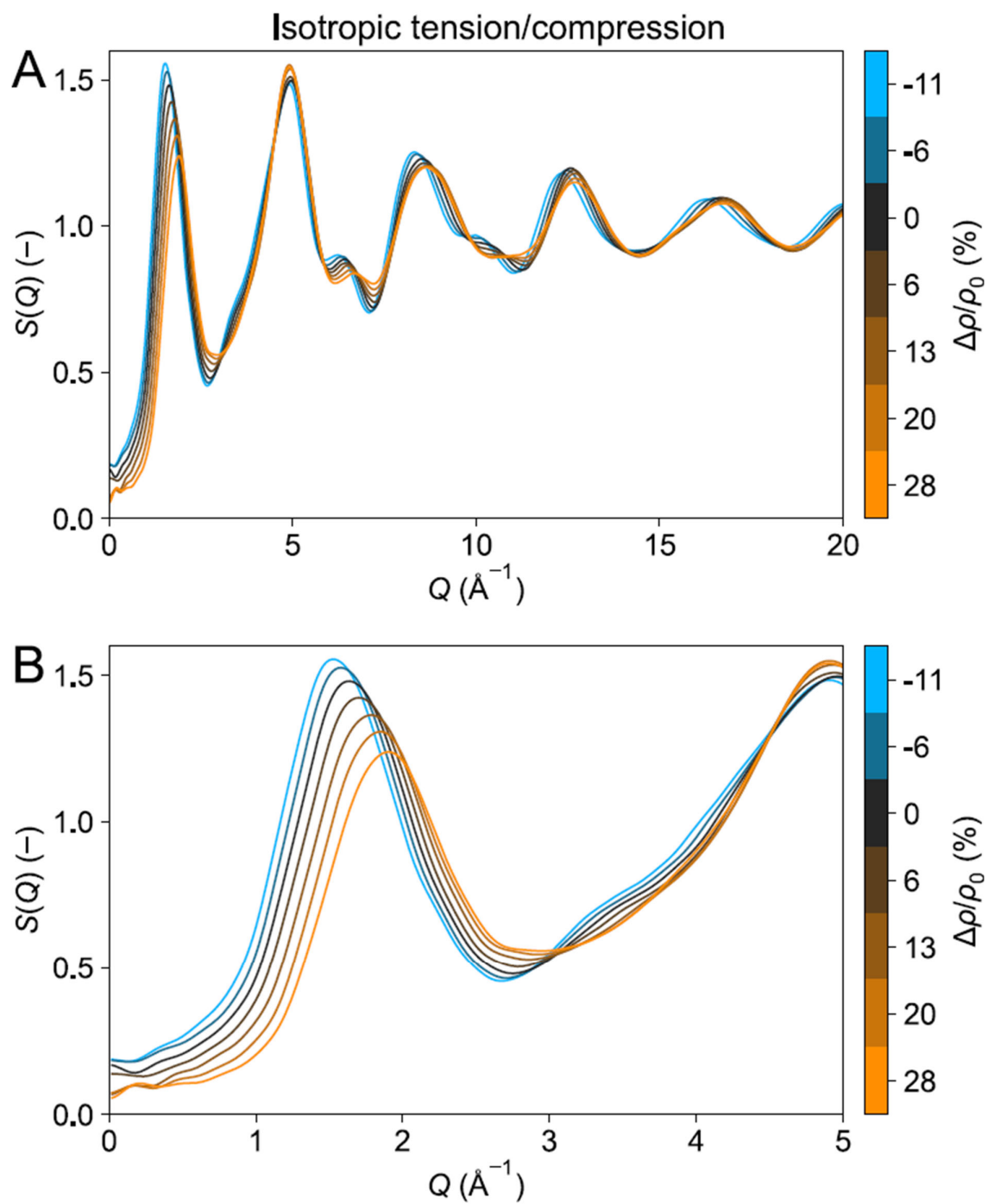

**Fig. S2.** Evolution of the X-ray structure factor $S(Q)$ as obtained from classical molecular dynamics simulation of vitreous silica upon isotropic tension or compression. $\Delta\rho/\rho_0$ is the relative change in density ($\rho$). Results are shown for scattering vector $Q$ from **(A)** 0 to 20 Å$^{-1}$ or **(B)** 0 to 5 Å$^{-1}$.

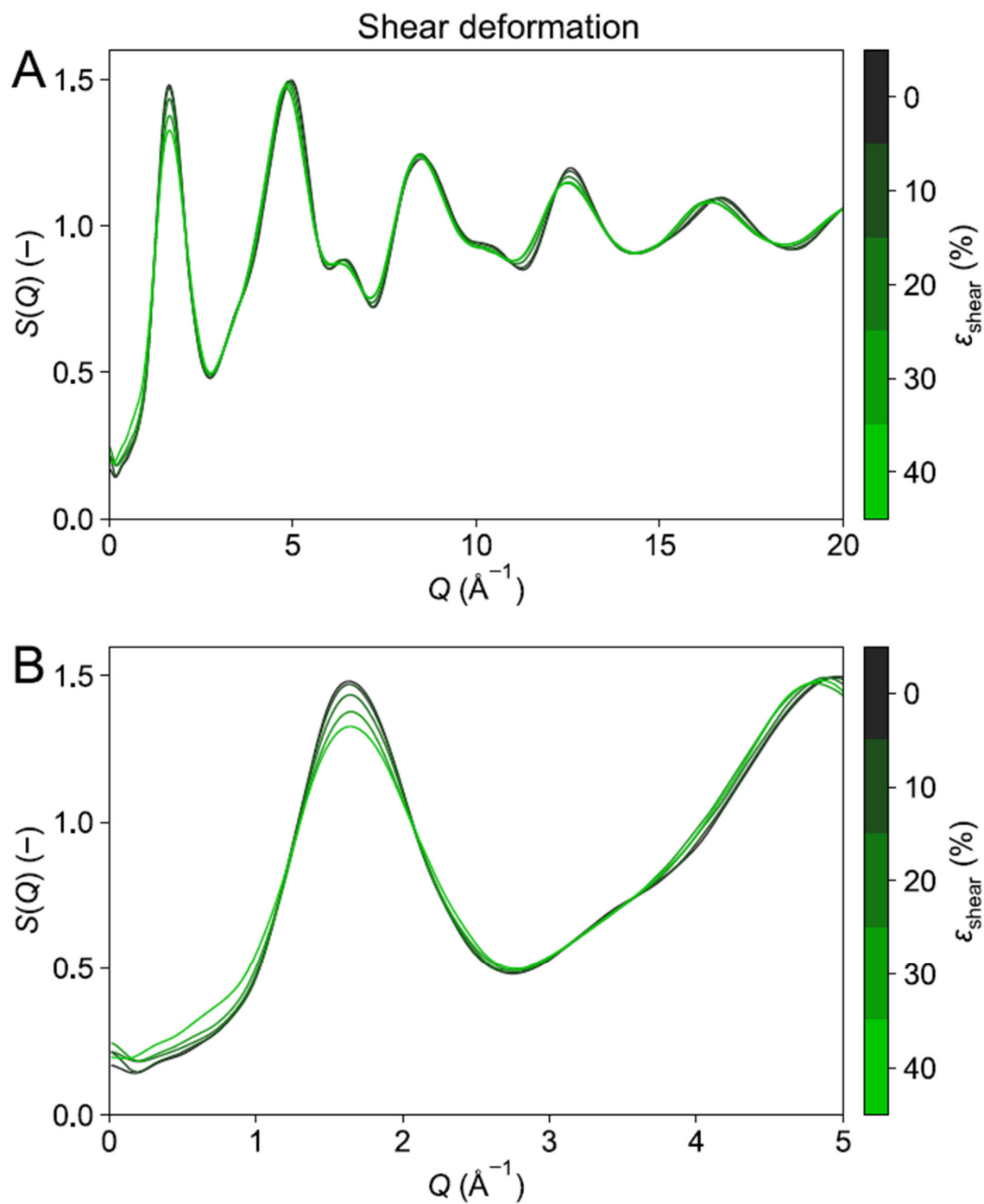

**Fig. S3.** Evolution of the X-ray structure factor $S(Q)$ as obtained from classical molecular dynamics simulation of vitreous silica upon shear deformation for shear strains ($\varepsilon_{strain}$) up to 40%. Results are shown for scattering vector $Q$ from **(A)** 0 to 20 Å$^{-1}$ or **(B)** 0 to 5 Å$^{-1}$.

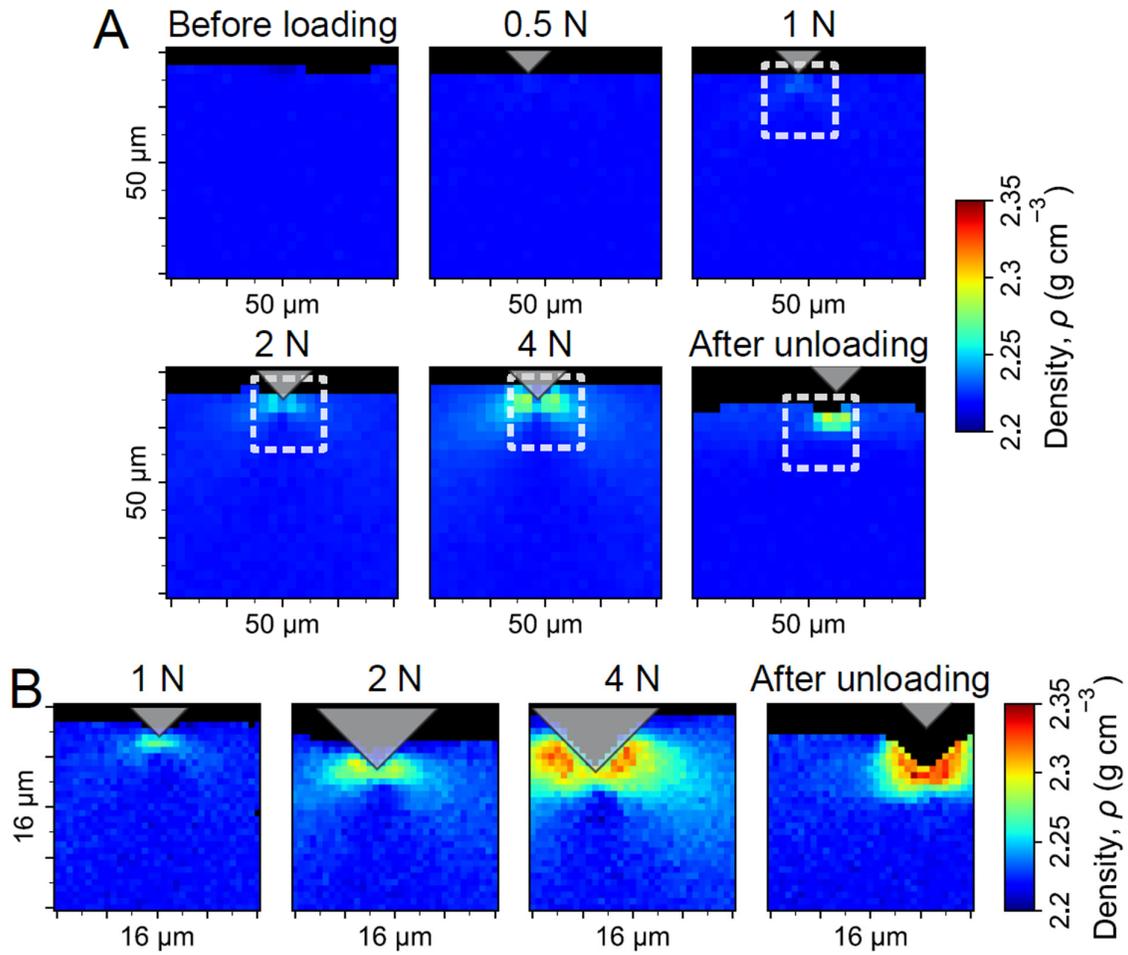

**Fig. S4.** Density heat maps, corresponding to Fig. 3 in the main text, but consistently using a color scale of 2.20 to 2.35 g cm$^{-3}$. Results are shown at different stages of the indentation cycle for **(A)** large-area scans of 2 μm resolution and **(B)** small-area scans of 0.5 μm resolution. The white-dashed marked boxes in panel A denote the areas that have been subjected to the additional scans shown in panel B. The grey triangles represent the position of the indenter, which is slightly shifted between the scans.

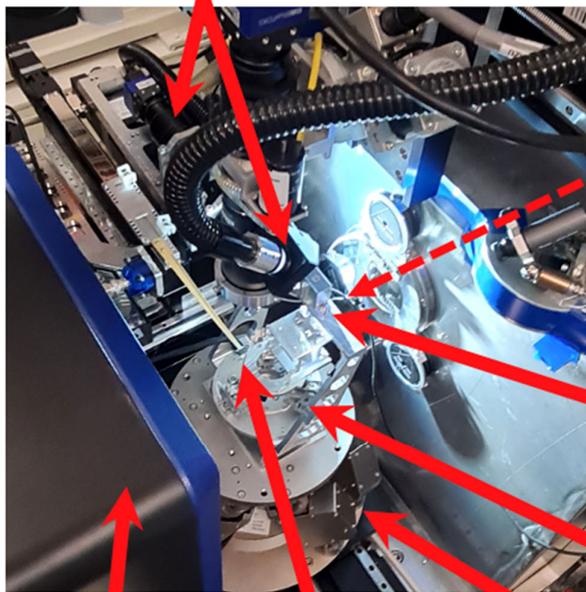

**Fig. S5.** Photo of the experimental setup at the beamline. Close-up photo of the indentation setup is shown in Fig. S6.

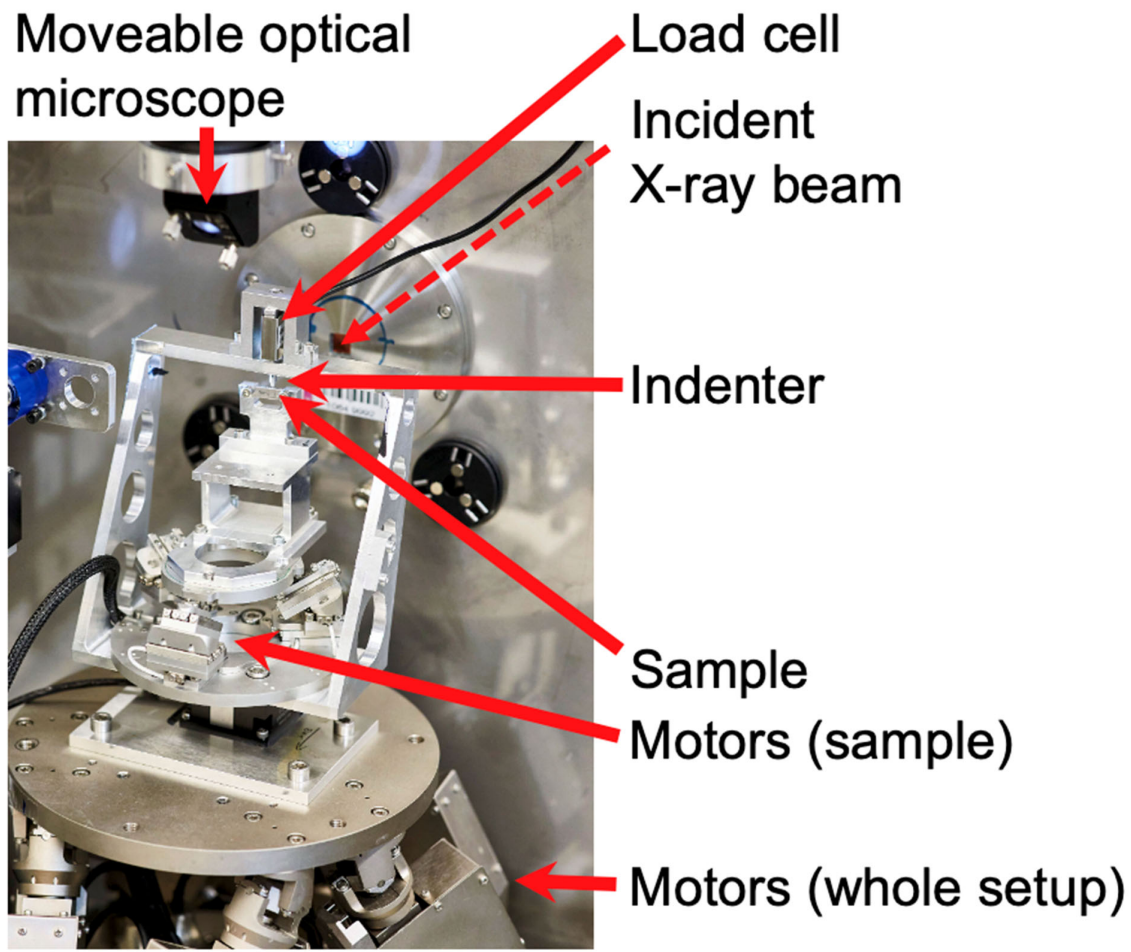

**Fig. S6.** Close-up photo of the indentation setup, highlighting the position of the load cell, indenter, and sample (not exact sample setup used for the present experiment).